# KRAB Algorithm - A Revised Algorithm for Incremental Call Graph Generation

Rajsekhara Babu *
School of Computing
Science and Engineering
VIT University
Vellore, India
*mrajasekharababu
@vit.ac.in*

Krishnakumar V.
School of Computing
Science and Engineering
VIT University
Vellore, India
*venkatasubramanian2011
@vit.ac.in*

George Abraham
School of Computing
Science and Engineering
VIT University
Vellore, India
*george.abraham2011
@vit.ac.in*

Kiransinh Borasia
School of Computing
Science and Engineering
VIT University
Vellore, India
*borasiakiransinh.ranjitsinh2011
@vit.ac.in*

**Abstract:** This paper is aimed to present the importance and implementation of an incremental call graph plugin. An algorithm is proposed for the call graph implementation which has better overall performance than the algorithm that has been proposed previously. In addition to this, the algorithm has been empirically proved to have excellent performance on recursive codes. The algorithm also readily checks for function skip and returns exceptions.

## I. INTRODUCTION

Call graph analysis is essential for understanding the execution of the program and also for debugging. The existing static call graph representation requires re-computation of the entire code whenever a small modification is made. An innovative approach is to follow an incremental approach, which would require the computation of the call graph of only the modified part and add it as an increment to the existing call graph. In this graph, each node represents a procedure and each edge from node A to node B indicates that procedure A has called procedure B. The main benefit of such call graphs is it provides a basic program analysis for human understanding. With both the static and dynamic call graphs, the programmer is able to understand the execution of his program and also aids him in debugging, when required. One simple application of call graphs is finding procedures that are never called. Call graphs can be dynamic or static. A dynamic call graph records the execution of a program and hence exactly describes one run of the program whereas a static call graph represents every possible run of the program. By tracking a call graph, it may be possible to detect anomalies of program execution or code injection attacks.

## II. RELATED WORK

There are a lot of techniques which are used for the generation of call graphs. They include Reachability Analysis [2][7], Inter Procedural Class Analysis [1], Class Hierarchy Analysis [3][7] and Fast Static Analysis [4]. Reachability is basically the ability of the graph algorithm to get from one vertex of the graph to another. A function A is said to be reachable from function B is there is a function call for A in the definition of function B. Reachability Analysis is performed by applying the algorithm recursively on each and every function whose function call is available from the entry point. Reachability Analysis can be improved by pre-computing the names and storing them in a data-structure (like the hash table) through which we can easily link the dependencies. In Object Oriented programming languages, the target procedure of a function call cannot be resolved just by examining the source code. This is due to the use of polymorphism present in the object oriented languages. The invoked function is strongly coupled with the object of the class which invokes it. Therefore for OOPL, inter-procedural data and flow analysis is important to understand the control flow of the functions. Figure 1 shows the class hierarchy in a real world scenario. Here, Class A acts as the base class while all the other are derived from Class A. This is called as the Inheritance graph. This graph helps the compiler to get an idea of how the transition occurs from class to class, thereby improving the run time performance of the call graph algorithm. Fast Static Analysis takes care of the virtual functions in the programs, which are extremely hard for the computer programs to analyze. Thereby, it removes a lot of overhead and improves the speed of the algorithm many times. It is used in object oriented programming languages since; it is more effective in them.

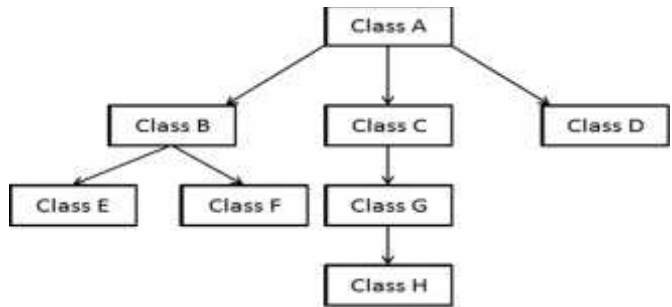
**Figure 1. Inheritance Graph**

### III. APPROACH

Eclipse is mainly used as an IDE for Java programs. Java being an Object Oriented Programming Language, Class Hierarchy Analysis is the approach which is most suitable for the Eclipse IDE plugin. Using Class Hierarchy Analysis, we can easily resolve the discrepancies that arise due to inheritance and polymorphism, thereby serving a dual purpose.

*Classical Approach*

Figure 2 represents the classic algorithm for call graph generation using the Class Hierarchy method. This algorithm is efficient for most of the cases, but has two serious drawbacks. Firstly, the performance reduces drastically when the system has recursive calls. The data structure which is used in this algorithm creates an instance for every recursive call that is made by the function. This puts added stress on the graph data-structure that we use thereby, increasing the amount of time taken for its execution. Secondly, once the function completes the execution, the graph has to backtrack to its parent. In the generic algorithm, the data structure doesn't take care of this issue and leaves it to the system. This results in more processor overhead and concurrently more execution time.

Since we are dealing with an incremental approach, it is necessary that we assign the pointer variables in the data structure components for easy traversals in case there are real-time changes that are made by the user in the structure of the program. The back-tracking becomes a time consuming task due to the absence of these pointers.

```
Classical Algorithm [5]
1: Build / Update Class Hierarchy
2: Find Program Entry Points (Add to
   work list)
3: Find Last Edited Method (Add to
   work list)
4: Pull method from head of list
5: Parse method for local declarations
6: Parse method for method calls
7: Find target methods corresponding to
   call site
8: Add Links between methods for calls
   and returns
9: Propagate live types across calls and
   returns
10: Find methods that need to be
    reprocessed and place back in work
    list
11: Output all connected components
```
**Figure 2. Classical Algorithm**

*KRAB Algorithm – Our Approach*

Figure 3 gives the pseudo code for the algorithm that we propose. This algorithm takes care of both the recursion related issue and the time taken due to back tracking. For this purpose, we make the graph data-structure using the help of a stack. The stack keeps a note of every function call encountered and automatically places a pointer from the current node to the parent node as a predecessor the moment the callee method returns its control to the caller method.

In this algorithm, we push every method call (in the form of a node) invoked by a specific object into a temporary stack. Now the control traverses the called function and checks for any method calls in it. It there are calls, it first checks whether it's a recursive call or a non-recursive call. If it is a recursive call the data-structure is pointed to itself rather than creating a new node instance. But, in order to verify the node traversal, every recursive call too is pushed into the stack. When the function returns its control, two activities are performed. First, a logical pointer link is generated from the current method to the parent method. This necessarily creates a predecessor pointer from the current method to the parent method. Second, the control is transferred from the current method to the parent method by popping the stack, since according to the algorithm; the node at the top of stack always has the control of the method.

```
KRAB Algorithm
1. Start.
2. If function is the start function (main),
   push in Stack and set it as root
   node.
3. While function call present
   3.1. Push into stack
   3.2. If call is recursive
      3.2.1. Loop node if its not
             already looped
   3.3. If call is not recursive
      3.3.1 Push into stack and
            set the node as leaf node
            with the above node
            as parent
   3.4. Traverse the function
   3.5. If function call present Goto 3
   3.6. Pop from stack.
   3.7. Set an in-predecessor from the
        current node to Top of Stack
        node
   3.8. Goto node indicating Top of
        Stack
4. If Top of Stack has start function
   Pop the stack
5. If Stack is not empty, throw exception
   Else Plot Graph and Exit
```

**Figure 3. Proposed Algorithm – KRAB Algorithm**

### IV. ANALYSIS

In the classical approach, we know that the time taken for building a standard class hierarchy takes on an average of $(n^2)$, since this is the time required for constructing a graph data structure. Finding the target methods (step 7) will take $(n)$ time. Finding the methods that needs to be reprocessed takes on an average $(n)$. The time complexity of searching is taken as $(n)$ since, we have to perform linear search as the given set of values may not be in a sorted manner. This gives the quadratic equation f(n) as:

$$f(n) = n^2 + n + n$$
$$= n^2 + 2n \quad \text{-------- [1]}$$

Hence, the minimum iterations that needs to be carried out in the classical algorithm is $n^2 + 2n$ iterations where n is the number of function calls present in a particular code.

For KRAB Algorithm, we can neglect the time required for push and pop operations in the stack and concentrate on the critical operations, which is the operations in the while loop.

Here, if we have n function calls, the while loop runs n times. In the best case scenario, there are no nested function calls, which give the Time Complexity as $(n)$ since the while loop runs once and ends. In the worst case, there would be nested function calls inside every function, which would give the Time Complexity as $O(n^2)$ since, the while loop has to execute n times within every function call due to nesting. On an average case, we can say that some function may have nested calls while some may not have nested calls. Let us assume that there are "log n" number of calls which have nested calls. This gives us the equation of k(n) as:

$$k(n) = n \log n \quad \text{(for average case)} \quad \text{-------- [2]}$$
$$= n^2 \quad \text{(for worst case)} \quad \text{-------- [3]}$$
$$= n \quad \text{(for best case)} \quad \text{-------- [4]}$$

Table 1, shows a selected sample input that we have taken for analyzing our algorithm with the classical algorithm. The first column represents the number of function calls that may occur in a particular code. The empirical analysis gives the results as given in the table. This data has been plotted in the form of bar graphs to avail the following output.

| Function Calls | f(n) | $k_w(n)$ | $k_a(n)$ | $k_b(n)$ |
|---|---|---|---|---|
| 200 | 40400 | 40000 | 1059.664 | 200 |
| 400 | 160800 | 160000 | 2396.587 | 400 |
| 600 | 361200 | 360000 | 3838.16 | 600 |
| 800 | 641600 | 640000 | 5347.693 | 800 |
| 1000 | 1002000 | 1000000 | 6907.76 | 1000 |

**Table 1. Performance Analysis**

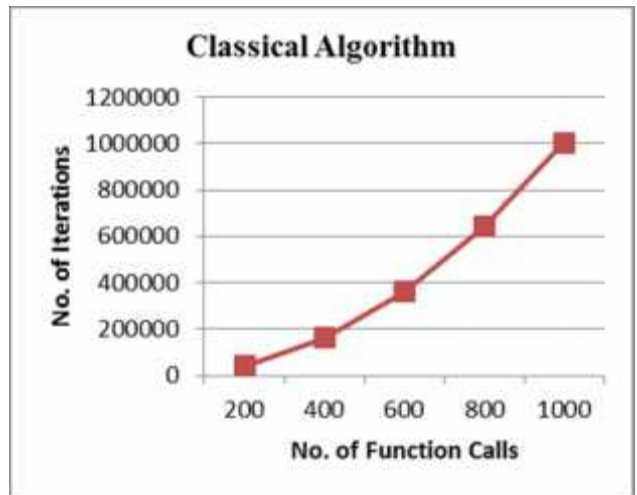

**Figure 4. Classical Algorithm**

As we can see in the Figure 4, the algorithm has a complexity of $O(n^2)$ in the best, worst and average case. Hence it is very inefficient for creating the call graphs.

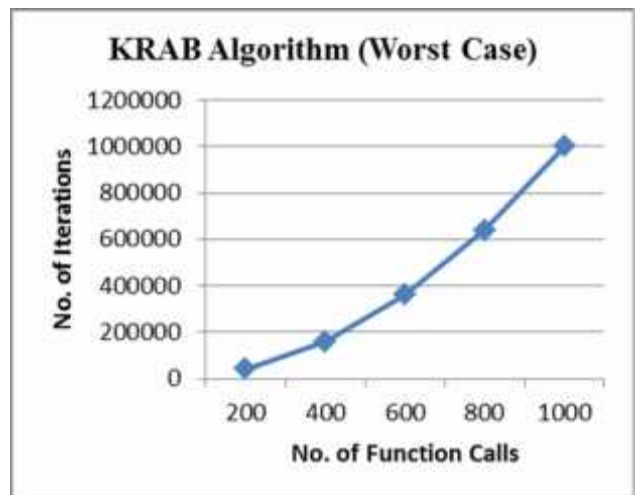

**Figure 5. KRAB Algorithm (Worst Case)**

The worst case for KRAB Algorithm occurs when each and every function call in the program has nested function calls within it. The worst case complexity is $O(n^2)$ which is the same as that of the average complexity of the classical algorithm.

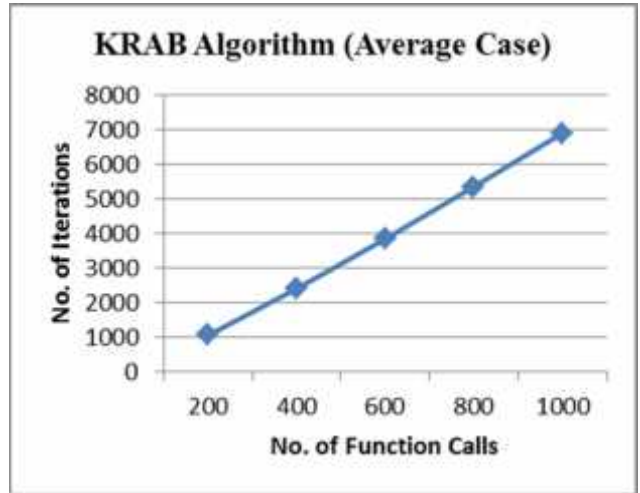
**Figure 6. KRAB Algorithm (Average Case)**

For the average case, we have assumed that some of the function calls are either recursive or non-nested while others have nested function calls. Under such cases, the algorithm exhibits comparatively better complexity. Suppose there are "log n" instances which have nested method calls out of a possible "n" method calls, then we will have a complexity of O(n log n). This is an excellent improvement over the classical algorithm as we can see through the graph. If we have 1000 function calls, it requires more than one million iterations for the classical algorithm, while it requires only 6900 iterations for the KRAB algorithm.

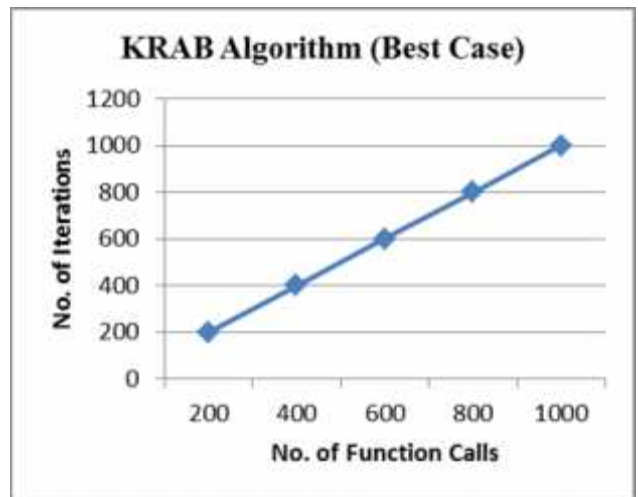
**Figure 7. KRAB Algorithm (Best Case)**

The best case assumes that every function call is either a recursive call or a non-nested function call. KRAB algorithm has a best case complexity of O(n) which is a huge improvement since the No. of iterations is equal to the No. of Function Calls required.

## V. CONCLUSION

In this paper we have proposed our algorithm for incremental call graph analysis and also analyzed its efficiency over the classical algorithm. The KRAB Algorithm has many advantages over the classical algorithms as well as some drawbacks. The usage of stacks in the KRAB algorithm makes it easy to keep track of the program controls. The time required to compute the predecessor of the current node is less and is equivalent to the time required for popping the stack. It works better and more efficiently on programs which has recursive method calls. The time complexity of the KRAB algorithm is better than the classical algorithm in all possible cases. It is capable of identifying any skips that may occur during the execution of the program. If it occurs, the stack won't be empty after the last pop operation, thereby, throwing an exception to the coder. Even though KRAB algorithm is more efficient that the classical algorithm, it requires an extra space complexity of 'n' during its first iteration for the temporary stack. The performance reduces when virtual functions are used in the programs. Another limitation of the KRAB algorithm is that it doesn't support Auto Boxing of the function calls on primitive types. Class Hierarchy technique requires the classes to be named. If anonymous inner classes are used in the code, then handling them becomes a tedious task.

## VI. FUTURE WORK

This algorithm is still in its infancy and requires huge number of improvements. This algorithm has been improved to extend the support for virtual functions and auto-boxing. The extension to virtual functions can be achieved by integrating the Hierarchical Analysis with Fast Static Analysis. Fast Static Analysis provides excellent results if virtual functions are present. Hence, a hybrid of these two Analytical methods will prove to be helpful in improving the stability of the algorithm. Auto Boxing is used by most of the programmers since it's easy to use and reduces the lines of code. Therefore, compatibility for Auto Boxing is an essential point that would be taken care of during the improvement.

## REFERENCES

[1] D. Grove and C. Chambers, "A Framework for Call Graph Construction Algorithms", ACM Transactions, ACM, Vol. 3, No 6, November 2001
[2] A. Srivastava, "Unreachable Procedures in Object-oriented Programming", WRL Research Report, August 1993
[3] J. Dean, D. Grove and C. Chambers, "Optimization of Object-Oriented Programs Using Static Class Hierarchy Analysis", 1994
[4] D. F. Bacon and P. F. Sweeney, "Fast Static Analysis of C++ Virtual Function Calls", ACM Conference, ACM, October 1996
[5] U. Ismail, "Incremental Call Graph Construction for the Eclipse IDE", University of Waterloo Technical Report, 2009
[6] W. Zhang and B. Ryder, "Constructing Accurate Application Call Graphs For Java To Model Library Callbacks"
[7] F. Tip and J. Palsberg, "Scalable Propagation-based Call Graph Construction Algorithms"